\documentstyle[aps,pra,psfig,epsfig,twocolumn,floats]{revtex}
\begin{document}

\draft
\title{
Conditional teleportation using optical squeezers and photon counting
}
\author{Jens Clausen,$^{1}$
Tom\'{a}\v{s} Opatrn\'{y},$^{1,2}$
and Dirk--Gunnar Welsch$^{1}$
 }
\address{
$^{1}$ Theoretisch-Physikalisches Institut,
Friedrich-Schiller-Universit\"at, Max-Wien-Platz 1, D-07743 Jena,
Germany \\
$^{2}$ Faculty of Science, Palack\'{y} University, 
Svobody 26, CZ-77146 Olomouc, Czechia \\
}
\date{\today}

\maketitle

\begin{abstract}

We suggest a scheme of conditional teleportation of quantum states of optical
fields using squeezers and photon counting. Alice feeds the mode whose state is
desired to be teleported and one mode of a two-mode squeezed vacuum into a
parametric amplifier and detects output photon numbers. The result is then
communicated to Bob who shifts the photon number of his part accordingly. We
show that for some classes  of states the method can yield, with reasonable
success probability, a  teleportation fidelity close to unity. The method is a
principally realizable modification of a recently  proposed scheme [G.J.
Milburn and S.L. Braunstein,  Phys. Rev. A {\bf 60}, 937 (1999)], where
measurements of the  photon-number difference and the phase sum are
considered. 

\end{abstract}


%%%%%%%%%%%%%%%%%%%%%%%%%%%%%%%%%%%%%%%%%%

\section{Introduction}

In quantum teleportation, an unknown state of a system is destroyed and 
created on another, distant system of the same type. The method was first 
suggested in \cite{Bennett} and realized in \cite{Zeilinger} for  discrete
variables, namely photonic (polarization) qubits. Subsequently, the  concept
has been extended to continuous variables \cite{Vaidman,Kimble}, and then
realized experimentally to teleport a coherent state by means of parametrically
entangled (squeezed) optical beams and quadrature-component measurements
\cite{Furusawa}.  The concept of teleportation of continuous quantum variables
has been further elaborated in \cite{Everybody}.

The basic requirement of quantum teleportation is  that the two parties share
an entangled state with each other. In continuous-variable teleportation of
quantum states of  optical field modes, a two-mode squeezed vacuum is suited
for playing the role of the entangled state. The quadrature components  $\hat
q_k$ and $\hat p_k$ ($[\hat q_k,\hat p_k]$ $\!=$ $\!i$, \mbox{$k$ $\!=$
$\!1,2$}) are correlated and anti-correlated, respectively, such that
%TOM
$\Delta(\hat q_1$ $\!-$ $\!\hat q_2)$  $\!<$ $\!1$ and
$\Delta (\hat p_1$ $\!+$ $\!\hat p_2)$
$\!<$ $\!1$.  For large squeezing, the correlations approach    the original
Einstein-Podolsky-Rosen (EPR)  correlations \cite{EPR} (for EPR correlations in
optical fields, see, e.g., \cite{Reid,Ou}).

The first scheme of teleportation that uses an optical two-mode  squeezed
vacuum is based on (single-event) quadrature-component  measurements exploiting
the above mentioned quadrature-component  correlations \cite{Kimble}.  Later
on, it has been realized that  there are photon-number and phase correlations
in  a two-mode squeezed vacuum which could also be used for a potential
teleportation protocol \cite{Milburn}.  In the scheme in \cite{Milburn} it is
assumed that a  measurement of the photon-number difference and
the phase sum of  the two modes on Alice's side is performed. The obtained
information is then sent to Bob who has to transform the  quantum state of his
mode by appropriate phase and photon-number shifting,  thus creating the
resulting teleported state. The scheme is conditional,  as for some measured
photon-number differences the state that is desired  to be teleported cannot be
re-created by Bob.

Unfortunately, the scheme in \cite{Milburn} requires  phase measurements for
which no methods have been known so far.  Is there any hope to realize
teleportation based on  such a scheme or a related one?  In this paper we
suggest a viable modification of the scheme  proposed in \cite{Milburn} which
is based on (single-event)  photon-number measurements on the output of a
parametric amplifier (squeezer). The scheme is also conditional, and it applies
to certain classes of quantum states. Even though the scheme is  not universal,
it can produce 
%(for the same degree of squeezing) 
for some states and some measurement events
higher  teleportation
fidelities than the scheme based on  quadrature-component measurements
\cite{Kimble}.

The paper is organized as follows. In Sec.~\ref{S-scheme}  we present the
scheme and derive the expression for the  teleported quantum state. In
Sec.~\ref{S-exper} we illustrate the method presenting numerical results, and
we  conclude in Sec.~\ref{S-concl}.

%%%%%%%%%%%%%%%  F I G U R E %%%%%%%%%%%%%%%%%%%%%%
\begin{figure}[!t!]
\noindent
\begin{center}
\epsfig{figure=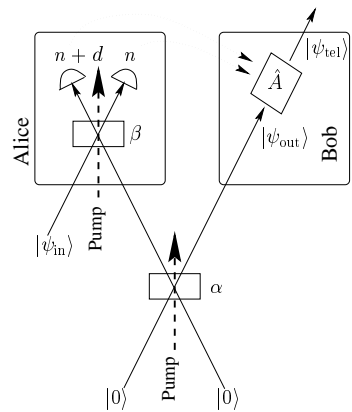,width=0.55\linewidth}
\end{center}
\caption{ 
Teleportation scheme as explained in the first paragraph
of Sec.~\protect\ref{S-scheme}.
}
\label{fig1}
\end{figure}
%%%%%%%%%%%%%%%%%%%%%%%%%%%%%%%%%%%%%%%%%%%%%%%%%%%%%%%%%

%%%%%%%%%%%%%%%%%%%%%%%%%%%%%%%%%%%%%%%%%%

\section{Theory}
\label{S-scheme}

Let us consider the scheme sketched in Fig.~\ref{fig1}.  The entangled state is
a two-mode squeezed vacuum produced by  the first parametric amplifier from the
vacuum state,  $\alpha$ being the squeezing parameter. One of the two output 
modes of the first parametric amplifier is then used as one of  the input modes
of the second parametric amplifier (squeezing parameter $\beta$),  and the mode
whose quantum state $|\psi_{\rm in}\rangle$ is desired to be teleported is the
other input mode.  Alice measures the photon numbers $n$ and $n$ $\!+$ $\!d$ 
($d$ $\!\ge$ $\!-n$) at the output of the second parametric  amplifier and
communicates the result to Bob. Owing to Alice's measurement, the state of the
mode that was sent to Bob from the first parametric amplifier has been
projected onto the state  $|\psi_{\rm out}\rangle$. Bob now reproduces the
input  state by means of the transformation \mbox{$\hat A|\psi_{\rm
out}\rangle$  $\!=$ $\!|\psi_{\rm tel}\rangle$}, where the operator $\hat A$
shifts the photon number according to  the measured photon-number difference
$d$.  

The three modes are initially (i.e., before they 
enter any of the parametric amplifiers) prepared in the states 
$|\psi_{{\rm in}}\rangle$, $|0\rangle$, and $|0\rangle$, where the 
state $|\psi_{{\rm in}}\rangle$ 
that is desired to be teleported can be written in the Fock basis as
\begin{equation}
 \label{1}
 |\psi_{{\rm in}}\rangle = \sum_{k} |k \rangle \langle k|\psi_{\rm in}\rangle.
\end{equation}
After passing the parametric amplifiers
and detecting $n$ and $n'$ photons in the outgoing modes (modes 
$0$ and $1$) on Alice's side, the state of Bob's mode (mode $2$) is
\begin{equation}
 \label{2}
 |\psi_{{\rm out}}\rangle_2 = P^{-\frac{1}{2}}
 {_0}\langle n | {_1}\!\langle n'| \hat S_{01}(\beta) \hat S_{12}(\alpha)
 |\psi_{{\rm in}}\rangle_0 |0\rangle_1 |0\rangle_2 ,
\end{equation}
where $P$ is the probability of that measurement event.
The two-mode squeeze operator $\hat S_{kl}(\alpha)$ is given by
\begin{eqnarray}
\label{3}
 \hat S_{kl}(\alpha) = \exp\!\left( \alpha^{*}\hat a_k  \hat a_l
 - \alpha \hat a_k^{\dag}  \hat a_l^{\dag}
 \right),
\end{eqnarray}
with $\hat a_k$ ($\hat a_k^{\dag}$) being the photon destruction
(creation) operator of the $k$th mode. It 
can be written in the Fock basis as
\begin{eqnarray}
  \label{4}
  \lefteqn{
  {_k}\langle m|\, {_l}\langle m'|  \hat S_{kl}(\alpha) |n\rangle_k |n'\rangle_l 
  =\delta _{m-m',n-n'} \,e^{i(m'-n')\varphi_{\alpha}}
  }
  \nonumber \\&&\hspace{2ex} \times \,
  (-1)^{n'}\sqrt{m! m'! n! n'!}\,
  \frac{(\sinh|\alpha|)^{n'} (\tanh|\alpha|)^{m'}}{(\cosh|\alpha|)^{n+1}} 
  \nonumber \\&&\hspace{2ex} \times 
  \sum_{j={\rm max}\{ 0,n'-n\} }^{{\rm min} \{ m',n' \} }
  \frac{\left( -\sinh ^2 |\alpha | \right)^{-j} }
  {j! (m'\!-\!j)!(n'\!-\!j)!(n\!-\!n'\!+\!j)!}\,,
\end{eqnarray}
where $\alpha$ = $|\alpha|e^{i\varphi_{\alpha}}$.
For the following it is useful to introduce the coefficients 
\begin{eqnarray}
 \label{5}
 \lefteqn{
 S^m_{m'}(d;\alpha) 
 = {_k}\langle m + d |\,{_l}\langle m|  \hat S_{kl}(\alpha) 
  |m'+d\rangle_k |m'\rangle_l
 } 
 \nonumber \\&&\hspace{9.5ex} 
  = \,{_k}\langle m |\,{_l}\langle m+d|  \hat S_{kl}(\alpha) 
  |m'\rangle_k |m'+d\rangle_l
 \nonumber \\&&\hspace{2ex}
  = e^{i(m-m')\varphi_{\alpha}} (-1)^{m'} 
 \sqrt{m!m'!(m\!+\!d)!(m'\!+\!d)!}
 \nonumber \\&&\hspace{2ex}\times\,
 \frac{(\tanh|\alpha|)^{m+m'}}{(\cosh|\alpha|)^{d+1}}
 \!\sum_{j=0}^{{\rm min} \{ m,m' \} }\! 
 \frac{\left( -\sinh ^2 |\alpha | \right)^{-j} }
 {j! (m\!-\!j)!(m'\!-\!j)!(d\!+\!j)!} \,.
 \nonumber \\&& 
\end{eqnarray}

The properties of the conditional quantum state $|\psi_{\rm out}\rangle$,
Eq.~(\ref{2}), in which the mode $2$ is prepared after the detection of 
$n$ and $n'$ photons in the modes $0$ and $1$ respectively, are 
qualitatively different for different sign of the observed difference 
$d$ $\!=$ $\!n'$ $\!-$ $\!n$. In the case when $d$ $\!\le$ $\!0$ is valid, 
then from Eq.~(\ref{2}) together with Eq.~(\ref{5}) it follows that 
($| \psi _{\rm out} \rangle_2$ $\!\to$ $\!| \psi _{\rm out} \rangle$)
\begin{equation}
\langle m | \psi _{\rm out} \rangle = P^{-\frac{1}{2}}  
   S^{n+d}_m(\!-d;\beta) S^m_0(0;\alpha) 
   \langle m\!-\!d |\psi_{\rm in}\rangle,
 \label{psim1}
\end{equation}
where the detection probability $P$ is given by
\begin{equation}
  \label{7}
  P = \sum_m  |S^{n+d}_m(-d;\beta)|^2
  |S^m_0(0;\alpha)|^2 
  |\langle m\!-\!d |\psi_{\rm in}\rangle|^2 .
\end{equation}
In the second case when $d$ $\!>$ $\!0$ is valid, we derive
\begin{equation}
 \langle m | \psi _{\rm out} \rangle \!=\!
 \left\{
 \begin{array}{l@{\ }l} 
 \!P^{-\frac{1}{2}}  S^n_{m-d}(d;\beta) S^m_0(0;\alpha) 
   \langle m\!-\!d |\psi_{\rm in}\rangle,
 & m \ge d, 
 \\[1ex]
 \!0, & m < d, 
 \end{array}
 \right.
 \label{psim2}
\end{equation}
where
\begin{equation}
 \label{9}
 P = \sum_{m\ge d}  |S^n_{m-d}(d;\beta)|^2
  |S^m_0(0;\alpha)|^2 
  |\langle m\!-\!d |\psi_{\rm in}\rangle|^2 .
\end{equation}

\begin{figure}[!t!]
\noindent
\begin{center}
\epsfig{figure=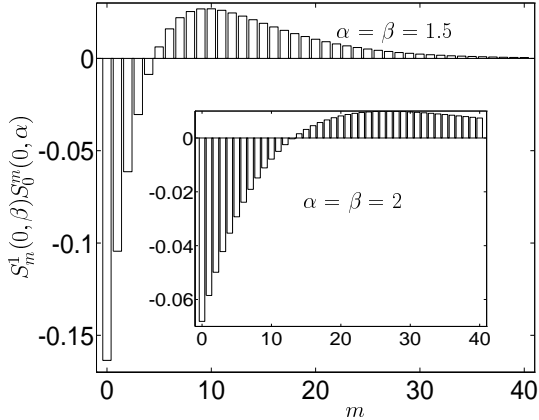,width=0.9\linewidth}
\end{center}
\caption{ 
The product
$S^{n+d}_{m}(d;\beta) S^m_0(0;\alpha)$ is shown for \mbox{$n$ $\!=$ $\!1$}
and \mbox{$d$ $\!=$ $\!0$}, and the squeezing parameters 
$\alpha$ $\!=$ $\!\beta$ $\!=$ $\!1.5$ and
$\alpha$ $\!=$ $\!\beta$ $\!=$ $\!2$.
}
\label{figss}
\end{figure}

{F}rom an inspection of Eqs.~(\ref{psim1}) and (\ref{psim2}) we see that when
the coefficients $S^{n+d}_m S^m_0$ and  $S^n_{m-d} S^m_0$, respectively,
change sufficiently slowly with $m$, then the state $|\psi_{\rm out}\rangle$,
Eq.~(\ref{2}), imitates the state  $|\psi_{\rm in}\rangle$, Eq.~(\ref{1}), but
with a  {\em shifted Fock-state expansion}, where the shift parameter  is just
given by the measured photon-number difference $d$. Obviously, if $d$ $\!<$
$\!0$ then the state $|\psi_{\rm out}\rangle$  does not contain any information
about the Fock-state expansion  coefficients $\langle m|\psi_{\rm
in}\rangle$ for  $m$ $\!<$ $\!|d|$. With regard to teleportation, this means
that  the method is conditional. Successful teleportation of a quantum state 
whose Fock-state expansion starts with the vacuum can only be achieved  if the
number of photons detected in the mode $1$ is not smaller than the number of
photons detected in the mode $0$. This limitation is exactly of the same kind
as in the scheme in \cite{Milburn}: the teleportation fidelity  tends sharply
to zero as the photon-number difference exceeds some (state-dependent)
threshold value.  Examples of the coefficients $S^{n+d}_{m} S^m_0$ are plotted in
Fig.~\ref{figss} for \mbox{$d$ $\!=$ $\!0$}.

%%%%%%%%%%%%%%%  F I G U R E %%%%%%%%%%%%%%%%%%%%%%
%%%%%%%%%%%%%%%%%%%%%%%%%%%%%%%%%%%%%%%%%%%%%%%%%%%%%%%%%

To complete the teleportation procedure, Bob transforms the 
state $|\psi_{\rm out}\rangle$ applying on it photon-number shifting. 
Thus, the teleported state is
\begin{equation}
 \label{10}
 |\psi_{\rm tel}\rangle 
 = \left\{ 
 \begin{array}{l@{\ \mbox{if}\ }l}
 \hat E ^{\dag -d} |\psi_{\rm out}\rangle & d < 0,
 \\[.5ex] 
 \hat E ^d  |\psi_{\rm out}\rangle & d > 0,
 \end{array} 
 \right.
\end{equation}
where 
\begin{equation}
 \label{11}
 \hat E = \sum_n |n\rangle \langle n+1|
\end{equation}
(i.e., the operator $\hat A$ in Fig.~\ref{fig1} is a power of $\hat E$ or $\hat
E^{\dag}$).
The teleportation fidelity is then given by 
\begin{equation}
 \label{12}
 F = |\langle\psi_{\rm in} | \psi_{\rm tel}\rangle |^2.
\end{equation}

For $d$ $\!\neq$ $\!0$, the teleportation scheme requires a  realization of the
transformations $\hat E$ and $\hat E ^{\dag }$.   Unfortunately, there has been
no exact implementation of these  transformations in quantum optics so far.
Photon adding and subtracting are transformations that are very close to  the
required ones. They are based on conditional measurement  and could be realized
using presently available experimental techniques \cite{Mohammed}. Their use of
course reduces the efficiency of the scheme. Thus, the scheme may be presently
confined to the case where  $d$ $\!=$ $\!0$.

%%%%%%%%%%%%%%%%%%%%%%%%%%%%%%%%%%%%%%%%%%%%%%%%%%%%%%%%%%%%%%%

\section{Results}
\label{S-exper}

{F}rom Eqs~(\ref{psim1}) and (\ref{psim2}) together with  Eqs.~(\ref{10}) --
(\ref{12}), the main results can be  summarized as follows. $(i)$ Fock states
can perfectly be teleported, i.e., the fidelity, Eq.~(\ref{12}), is equal to
unity, which follows from the fact that   parametric amplifiers conserve the
photon-number difference. Therefore,  high teleportation fidelities can also be
expected for  states with small photon number dispersion. For such states our 
method may be more suitable than the method in \cite{Kimble}, where 
teleportation via measurement of conjugate quadrature components is realized. 
On the other hand, high teleportation fidelities are not expected  for states
with large mean photon number and large photon-number dispersion.  In
particular, for teleportation of highly excited coherent states or phase 
squeezed states the method in  \cite{Kimble} may be more suitable.  $(ii)$ In
comparison to the method in \cite{Milburn}, our scheme does not require phase 
shifting of the output state $|\psi_{\rm out}\rangle$. The squeezing 
parameters $\alpha$ and $\beta$ can be chosen such that the coefficients 
$S_m^{n+d} S^m_0$ and $S_{m-d}^n S^m_0$ in Eqs.~(\ref{psim1})  and
(\ref{psim2}), respectively, are real, so that the Fock-state  expansion
coefficients $\langle m|\psi_{\rm out}\rangle$ have the same phase as the
coefficients  $\langle m - d|\psi_{\rm in}\rangle$.  $(iii)$ A high
teleportation fidelity can be expected, provided   that the values of the
coefficients $S_m^{n+d} S^m_0$ and $S_{m-d}^n S^m_0$ vary sufficiently slowly
with $m$ in the relevant range of the Fock-state expansion of the input state 
$|\psi_{\rm in}\rangle$. 
On the other hand, in ranges where the  coefficients change
rapidly, reliable teleportation cannot be achieved.  {F}rom Fig.~\ref{figss} it
is seen that   the $m$-range in which $S_{m-d}^n S^m_0$ slowly varies with $m$
increases with the strength of squeezing, and thus the class of states that can
be teleported reliably extends. 

%%%%%%%%%%%%%%%  F I G U R E %%%%%%%%%%%%%%%%%%%%%%
\begin{figure}[htb]
\begin{center}
\epsfig{figure=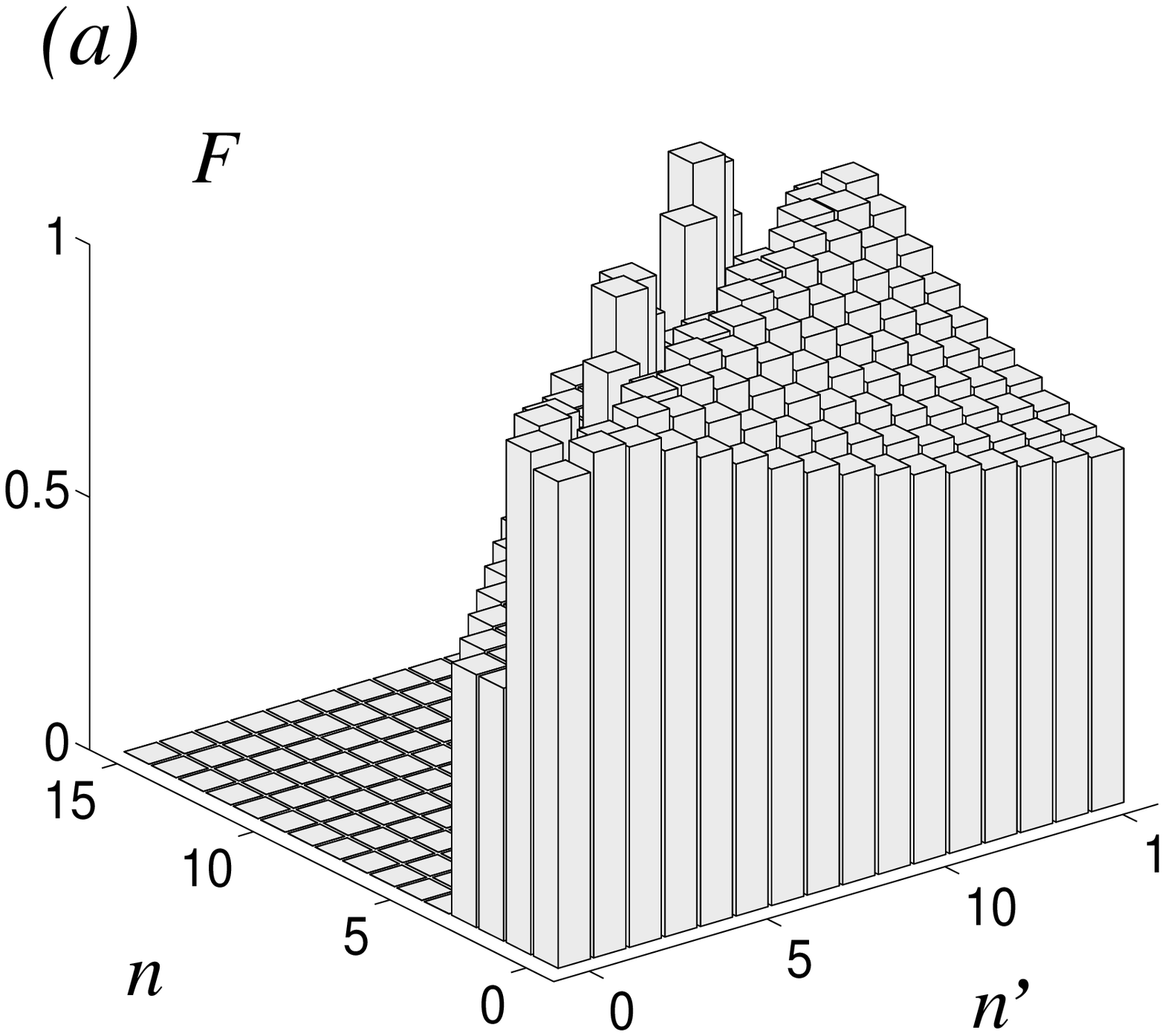,width=0.85\linewidth}
\epsfig{figure=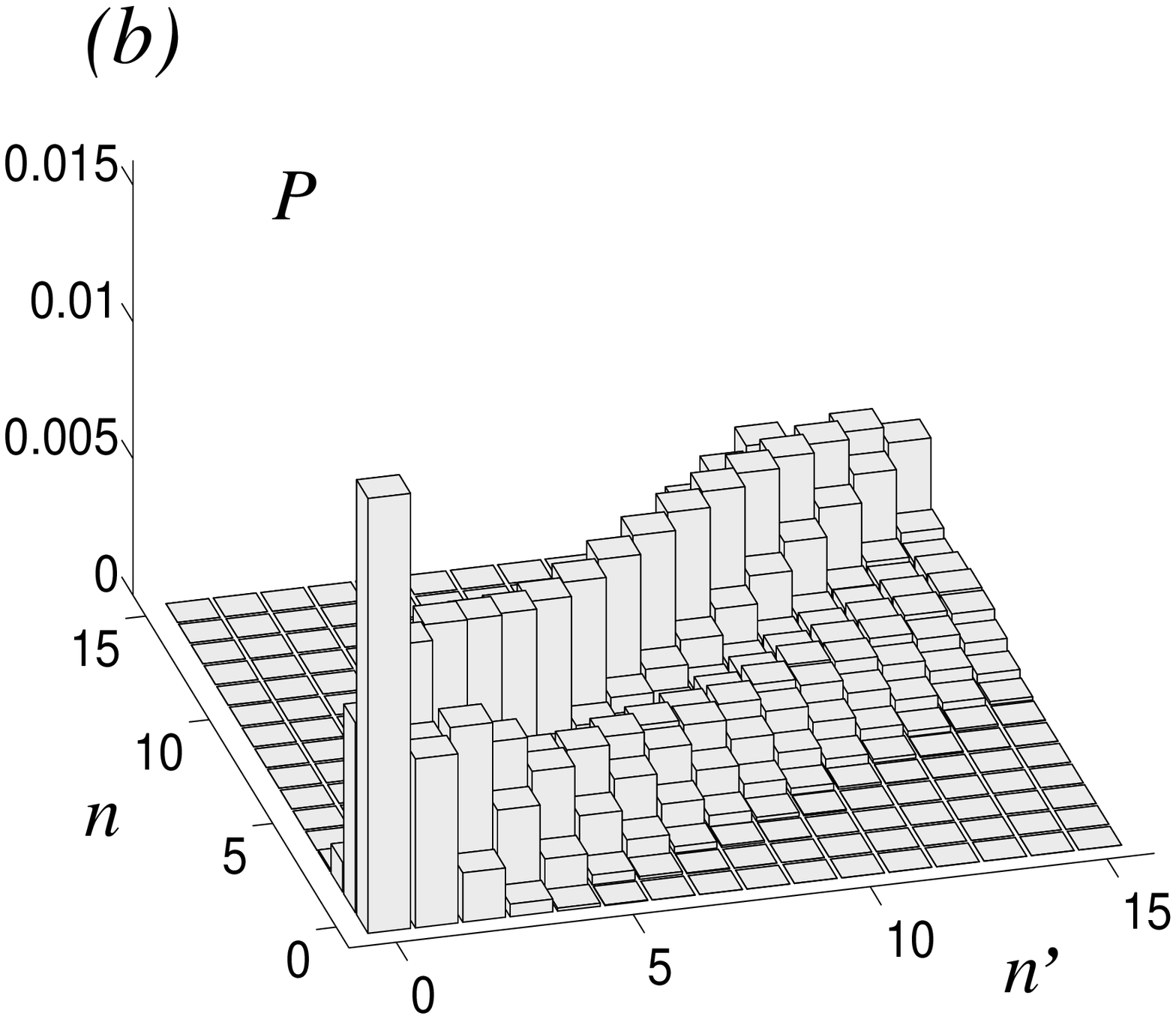,width=0.85\linewidth}
\end{center}
\caption{ 
Teleportation fidelity  $F$ $(a)$ and 
success probability $P$ $(b)$
in dependence on the measured photon numbers $n$ and $n'$.
The input state is $2^{-1/2}(|1\rangle + i|3\rangle)$ and 
the squeezing parameters are $\alpha$ $=$ $\beta$ $=$ 1.5.
}
\label{figfidels}
\end{figure}
%%%%%%%%%%%%%%%%%%%%%%%%%%%%%%%%%%%%%%%%%%%%%%%%%%%

In order to illustrate the method, we have calculated the teleported state,
assuming that input state is a superposition of two Fock states, $|\psi_{\rm
in}\rangle$ $\!=$ $\!2^{-1/2}(|1\rangle$ $\!+$ $\!i|3\rangle)$ and equal
squeezing parameters $\alpha$ and $\beta$ are used.  \mbox{Figure
\ref{figfidels}} presents the dependence on the detected photon numbers $n,n'$ 
of the teleportation fidelity $F$, Eq.~(\ref{12}), and the  success probability
$P$, Eqs.~(\ref{7}) and (\ref{9}).  We observe that close to the diagonal (but
not always directly on it) the fidelity reaches high  values close to unity. 
For the values   of $n,n'$ with $n$ $\!=$ $\!n'+2$ and $n$ $\!=$ $\!n'+3$ the
fidelity is exactly $0.5$, which indicates that the Fock state $|3\rangle$ was
in the input of the second squeezer and   has therefore been re-created in the
teleportation. For the values  of  $n,n'$ with $n$ $\!>$ $\!n'+3$ the fidelity
drops to zero and  so does the probability: such events do not occur for the
input state under consideration. 

To quantify the performance of the method, we 
have calculated the probability of events which yield teleportation 
fidelities larger than or equal to some upper value $F_{\rm u}$,
\begin{equation}
 P_{\rm u} = \sum_{n,n' \atop F(n,n') \ge F_{\rm u} } 
 P(n,n') ,
\label{CONPROB}
\end{equation}
where $P(n,n')$ is the success probability of detecting $n$
and $n'$ photons [Eqs.~(\ref{7}) and (\ref{9})], and
$F(n,n')$ is the corresponding teleportation fidelity.
In the example, we find that $P_{\rm u}$ $\!\approx$ $\!33\%$
for $F_{\rm u}$ $\!=$ $\!90\%$.
Measuring (in place of $n$ and $n'$ in 
our scheme) the quadrature components $X_0$ and $P_1$ in the
scheme in \cite{Kimble} would yield (for the same input state 
and the same squeezing parameter of the entangled state) 
\mbox{$P_{\rm u}$ $\!\approx$ $\!23\%$}.

\begin{figure}[!t!]
\begin{center}
\epsfig{figure=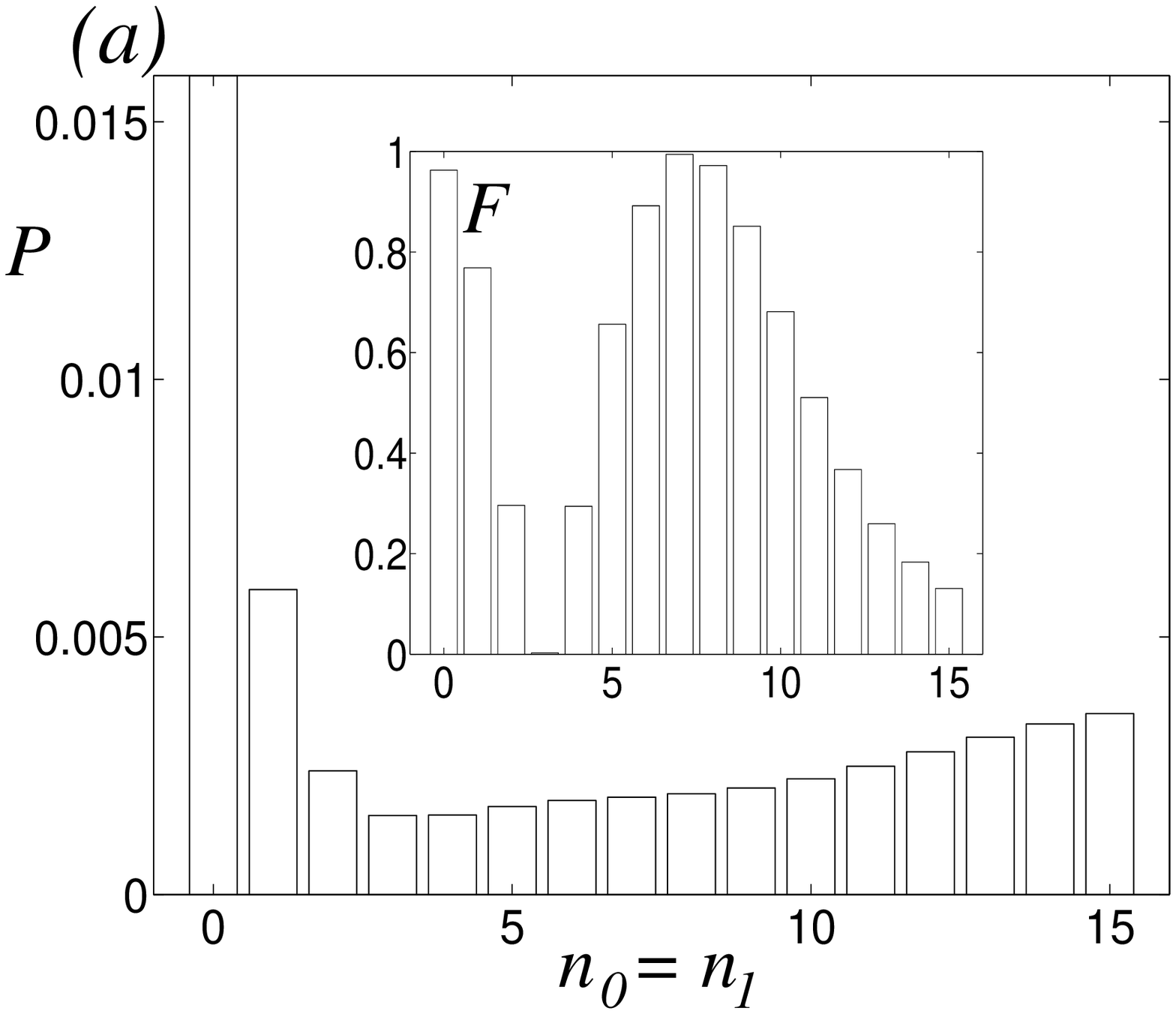,width=0.85\linewidth}
\epsfig{figure=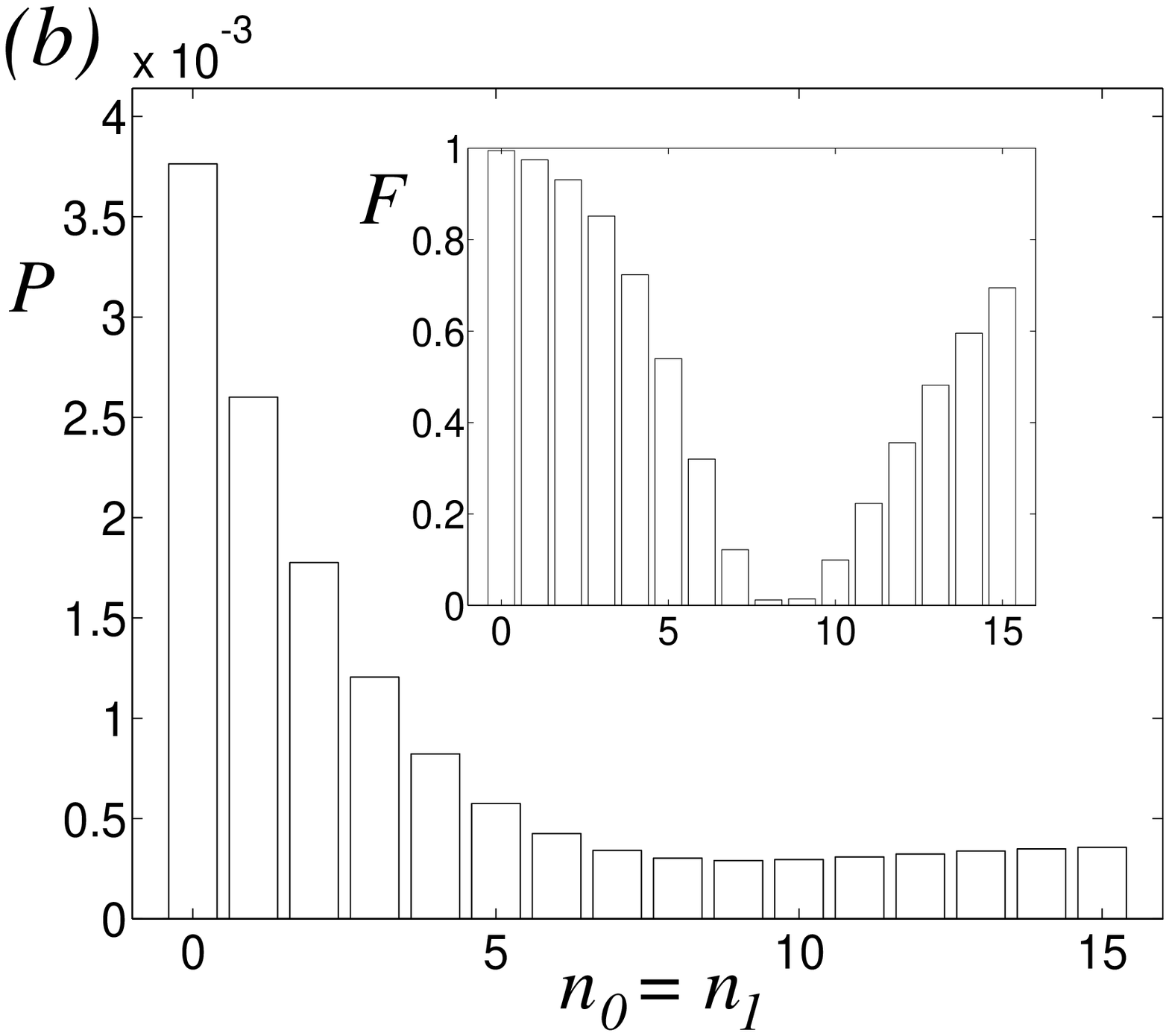,width=0.85\linewidth}
\end{center}
\caption{ 
Teleportation fidelity  $F$  and 
success probability $P$ 
in dependence on the measured photon numbers 
$n$ $\!=$ $\!n'$.
The input state is $2^{-1/2}(|1\rangle + i|3\rangle)$ and the squeezing
parameters are $\alpha$ $=$ $\beta$ $=$ 1.5 $(a)$, and
 $\alpha$ $=$ $\beta$ $=$ 2 $(b)$.
}
\label{figfidels2}
\end{figure}

Let us consider the more realistic
case where $n$ $\!=$ $\!n'$,  so that no
photon-number shifting is necessary.  {F}rom Fig.~\ref{figfidels2} we see that
with increasing strength  of squeezing a higher fidelity can be realized.
However, the  corresponding success probability decreases. In the figure, the
overall probability of realizing a fidelity higher than 90\% is $P$ $\!\approx$
$1.97\% $ for the squeezing parameters  $\alpha$ $\!=$ $\!\beta$ $\!=$
$\!1.5$,  whereas for $\alpha$ $\!=$ $\!\beta$ $\!=$ $\!2$ the probability
reduces to  \mbox{$P$ $\!\approx$ $\!1.06\% $}. 

Clearly, if the input states are completely unknown, it cannot be predicted
with what fidelity a state would be teleported. The scheme  applies if the
input states can be confined to a certain class of states, so that from an
estimated fidelity it can  be decided which photodetection results would
represent a successful teleportation.

%%%%%%%%%%%%%%%  F I G U R E %%%%%%%%%%%%%%%%%%%%%%
%%%%%%%%%%%%%%%%%%%%%%%%%%%%%%%%%%%%%%%%%%%%%%%%%%%

%%%%%%%%%%%%%%%%%%%%%%%%%%%%%%%%%%%%%%

\section{Summary and Conclusions}
\label{S-concl}

We have suggested a viable modification of the teleportation  scheme proposed
in \cite{Milburn}.  Our scheme avoids the phase sum measurement that is not 
realizable at present. It uses instead  the property of a nondegenerate
parametric amplifier that the photon-number difference  of the output beams is
equal to that of the input beams.  However, the price of avoiding phase
measurements is a relatively low success probability of teleportation.

Our method and the method in  \cite{Kimble}, which is based on
quadrature-component measurements, may complement one another.  So, our method
is better suited to teleportation of states with small photon number
dispersion  (Fock states can be teleported with fidelity equal to unity in
principle). The  method in \cite{Kimble} is more suitable for teleportation of
states with smooth  quadrature-component distributions.

Although our method is realizable in principle, there are several  non-trivial
experimental challenges. First, precise photodetection is  needed, i.e.,
detectors are required that are able to distinguish between different photon
numbers.  This does not only concern Alice's measurement but also Bob's
photon-number shifting, e.g., by means of photon adding  and subtracting.
Second, the photodetection should be sufficiently  mode-selective, i.e, one
must be able to distinguish whether  an incident photon comes from the mode
under study or from another part of the spectrum generated by the parametric
amplifiers.  A central problem in any scheme that exploits quantum coherence
is that of decoherence due to unavoidable losses.  The effect of decoherence
may be reduced, if the squeezing strengths are reduced. However, using smaller
squeezing decreases the available teleportation fidelity, so that one has to
find an  optimum regime for the teleportation of a given class of states, the
losses in the scheme, and the required fidelity.

%%%%%%%%%%%%%%%%%%%%%%%%%%%%%
\acknowledgments

This work was supported by the Deutsche Forschungsgemeinschaft.
We thank S. Braunstein and P. Kok for useful comments.

%%%%%%%%%%%%%%%%%%%%%%%%%%%%%%%%%%%%%%%%%%%%%%%%%%%%%%%%%%%%%%


\begin{references}

\bibitem{Bennett}
Ch. Bennett
et al.,
%TELEPORTING AN UNKNOWN QUANTUM STATE VIA DUAL CLASSICAL AND
%EINSTEIN-PODOLSKY-ROSEN CHANNELS 
Phys. Rev. Lett. {\bf 70}, 1895 (1993).

\bibitem{Zeilinger}
D. Bouwmeester, J.-W. Pan, M. Daniel, H. Weinfurter, and A. Zeilinger, 
Nature {\bf 390}, 575 (1997);
D. Boschi, S. Branca, F. DeMartini, L. Hardy, and S. Popescu, 
Phys. Rev. Lett. {80}, 1121 (1998).

\bibitem{Vaidman}
L. Vaidman,
% TELEPORTATION OF QUANTUM STATES (continuous variables included)
Phys. Rev. A {\bf 49}, 1473 (1994);

\bibitem{Kimble}
S.L. Braunstein and H.J. Kimble, Phys. Rev. Lett. {\bf 80}, 869 (1998).
%TOM
%T.C. Ralph and P.K. Lam, 
%Phys. Rev. Lett. 
%ibid {\bf 81}, 5668 (1998).

\bibitem{Furusawa}
A. Furusawa, J.L. S{\o}rensen, S.L. Braunstein, C.A. Fuchs, H.J. Kimble, and
E.S. Polzik, Science {\bf 282}, 706 (1998).

%DGW
\bibitem{Everybody}
T.C. Ralph and P.K. Lam,  Phys. Rev. Lett. {\bf 81}, 5668 (1998);
T. Opatrn\'{y}, G. Kurizki, and D.-G. Welsch, 
Phys. Rev. A {\bf 61}, 032302  (2000);
S. L. Braunstein, G. M. D'Ariano, G. J. Milburn, and M. F. Sacchi,
Phys. Rev. Lett. {\bf 84}, 3486 (2000).

\bibitem{EPR}
A. Einstein, B. Podolsky, and N. Rosen, 
Phys. Rev. {\bf 47}, 777 (1935).

\bibitem{Reid}
M.D. Reid and P.D. Drummond,
%QUANTUM CORRELATIONS OF PHASE IN NONDEGENERATE PARAMETRIC OSCILLATION
Phys. Rev. Lett. {\bf 60}, 2731 (1988);
M.D. Reid,
%DEMONSTRATION OF THE EINSTEIN-PODOLSKY-ROSEN PARADOX USING NONDEGENERATE
%PARAMETRIC AMPLIFICATION  
Phys. Rev. A {\bf 40}, 913 (1989).

\bibitem{Ou}
Z.Y. Ou, S.F. Pereira, H.J. Kimble, and K.C. Peng,
% Realization of the EPR paradox for continuous variables
Phys. Rev. Lett. {\bf 68}, 3663 (1992);
Z.Y. Ou, S.F. Pereira, and H.J. Kimble, 
%REALIZATION OF THE EINSTEIN-PODOLSKY-ROSEN PARADOX FOR CONTINUOUS-VARIABLES IN
%NONDEGENERATE PARAMETRIC AMPLIFICATION 
Appl. Phys. B {\bf 55}, 265 (1992).


\bibitem{Milburn}
G.J. Milburn and S.L. Braunstein, Phys. Rev. A {\bf 60}, 937 (1999).

\bibitem{Mohammed}
M. Dakna,  L. Kn\"oll, and D.-G. Welsch,
Euro. Phys. J. D {\bf 3}, 295 (1998);
M. Dakna, J. Clausen, L. Kn\"oll, and D.-G. Welsch,
Phys. Rev. A {\bf 59}, 1658 (1999), 
Phys. Rev. A {\bf 60}, 726(E) (1999);
J. Clausen, M. Dakna, L. Kn\"oll, and D.-G. Welsch,
Acta Phys. Slov. {\bf 49}, 653 (1999).
%[No. 4: Proceedings of the 6th Central-European Workshop on Quantum
%Optics, Chudobin, Czech  Republic, April 30 -- May 3, 1999].


\end{references}
\end{document}